\documentclass[10pt]{iopart}
\usepackage{iopams}
\usepackage{hyperref}
\hypersetup{
colorlinks=true,
linkcolor=blue,
citecolor=blue,
urlcolor=blue}
\newcommand{\intprod}{\mbox{$\;
\put(0,0){\line(1,0){.9}}\put(.9,0){\line(0,1){1.6}} \; \, \, $}}
\newcommand{\fr}{f(\mathcal{R})}

\begin{document}

\title[The gauge symmetries of $f(\mathcal{R})$ gravity with torsion in the Cartan formalism]
{The gauge symmetries of $f(\mathcal{R})$ gravity with torsion in the Cartan formalism}

\author{Merced Montesinos$^{1}$, Rodrigo Romero$^1$ and Diego Gonzalez$^{2}$}

\address{$^1$ Departamento de F\'isica, Cinvestav, Avenida Instituto Polit\'ecnico Nacional 2508, San Pedro Zacatenco, 07360, Gustavo A. Madero, Ciudad de M\'exico, M\'exico.}
\address{$^2$ Instituto de Ciencias Nucleares, Universidad Nacional Aut\'onoma de M\'exico, Apartado Postal 70-543, Ciudad de M\'exico, 04510, M\'exico}

\eads{\mailto{merced@fis.cinvestav.mx}, \mailto{rromero@fis.cinvestav.mx} and \mailto{diego.gonzalez@correo.nucleares.unam.mx}}


\begin{abstract}

First--order general relativity in $n$ dimensions ($n \geq 3$) has an internal gauge symmetry that is the higher--dimensional generalization of three--dimensional local translations. We report the extension of this symmetry for $n$--dimensional $\fr$ gravity with torsion in the Cartan formalism. The new symmetry arises from the direct application of the converse of Noether's second theorem to the action principle of $\fr$ gravity with torsion. We show that infinitesimal diffeomorphisms can be written as a linear combination of the new internal gauge symmetry, local Lorentz transformations, and terms proportional to the variational derivatives of the $\fr$ action. It means that the new internal symmetry together with local Lorentz transformations can be used to describe the full gauge symmetry of $\fr$ gravity with torsion, and thus diffeomorphisms become a derived symmetry in this setting.

\end{abstract}


\section{Introduction}

Despite the success of general relativity, the interest in theories beyond it, generically known as `modified gravity' theories has grown substantially in recent decades. In essence, the so-called modified gravity attempts to give an explanation to some cosmological and astrophysical observations that apparently do not fit in the theoretical framework of general relativity or matter fields coupled to general relativity, among them: accelerated expansion of the universe, the rotation curves of particles surrounding galaxies or the dynamics of galaxies in clusters, the large-scale structure of the universe, etc.~\cite{Sotiriou2010,Felice2010}.

Among the variety of modified gravity theories, one of the most straightforward generalizations of general relativity is  $\fr$ gravity.  In such theories, the Lagrangian is proportional to an arbitrary function~$\fr$ of the Ricci scalar~${\mathcal R}$, instead of just being linear in ${\mathcal R}$ as in general relativity.  There are three versions of $\fr$ gravity: metric $\fr$ gravity, Palatini $\fr$ gravity, and $\fr$ gravity with torsion. In the first case, the Lagrangian depends on the metric tensor only, because the spacetime connection is the Levi--Civita connection constructed out of the metric~\cite{Sotiriou2010}.  In the second case, it is assumed that the fundamental variables of the theory are the metric tensor and a torsion--free connection (see~\cite{Sotiriou2006, Felice2010}). In the third case, the fundamental variables of $\fr$ gravity with torsion are taken to be  either the metric tensor and a metric--compatible connection in the metric--affine formalism~\cite{Capozziello2007, Capozziello2008Jbundles, Capozziello2010}, or an orthonormal frame of $1$--forms and a metric--compatible connection in the Cartan formalism~\cite{Rubilar_1998}. 

On the other hand, it is well known that $\fr$ gravity with torsion in the Cartan formalism is by construction invariant under local Lorentz transformations and diffeomorphisms. These symmetries have been adopted for many years as the fundamental symmetries underlying the gravitational theories. Nevertheless, this paradigm has been recently challenged by a series of works showing that different equivalent sets of symmetries, which do not consider diffeomorphisms as fundamental, naturally emerge through the implementation of the converse of Noether's second theorem~\cite{Montesinos2017, MontesinosMatter2018, Montesinos2018Lovelock}. For instance, in Ref.~\cite{Montesinos2017} the symmetries of the $n$--dimensional Einstein--Cartan action with a cosmological constant ($n \geq 3$) are reformulated in this approach. It is shown there that the full gauge invariance of this action can be described by local Lorentz transformations and an internal gauge symmetry that is the higher--dimensional generalization of three--dimensional (3D) local translations. In this framework, infinitesimal diffeomorphisms are no longer regarded as fundamental but as a derived symmetry. Bearing in mind that the idea of replacing diffeomorphisms with 3D local translations has been useful to attack the problem of quantizing gravity in the 3D setting~\cite{Achucarro1986, Witten1988}, and given the advantages of $\fr$ gravity with torsion in comparison with other models, in the present paper we want to extend the analog of 3D local translations to the case of $\fr$ gravity with torsion. Furthermore, such internal gauge symmetry, along with local Lorentz transformations, would render $\fr$ gravity with torsion closer to ordinary gauge theories.

In light of this, here we show that there exist a new internal gauge symmetry for $n$--dimensional $\fr$ gravity with torsion in the Cartan formalism, that is the natural extension of the internal gauge symmetry reported in Ref.~\cite{Montesinos2017} for $n$--dimensional general relativity. In the case of $n$--dimensional  general relativity, {\it i.e.} $\fr={\mathcal R}-2\Lambda$, the new internal gauge symmetry collapses {\it off--shell} to the symmetry obtained in Ref.~\cite{Montesinos2017}. Furthermore, for a general $\fr$ theory, we find that infinitesimal diffeomorphisms can be written as a linear combination of local Lorentz transformations, plus the new internal gauge symmetry, plus terms proportional to the variational derivatives of the $\fr$ action. Thus, the new symmetry together with local Lorentz transformations can be taken as a set of fundamental symmetries to capture the full gauge invariance of $\fr$ theories of gravity with torsion. In this framework, diffeomorphisms are regarded as a derived symmetry. The new symmetry is obtained by applying the converse of Noether's second theorem, which involves the construction of a  non--trivial Noether identity. We achieve this by following an approach analogous to that used in Refs.~\cite{Montesinos2017, MontesinosMatter2018,Montesinos2018Lovelock}. An interesting property of the new internal gauge symmetry is that it depends explicitly on the spacetime dimension and the particular form of the function $\fr$, this in contrast to diffeomorphisms and local Lorentz transformations, which take the same structure independently of the form of the action from which they are deduced. Finally, we consider the case $\fr = \mathcal{R}^{n/2}$ ($n \geq 3$) and find, by using the converse of Noether's second theorem, that the corresponding action principle has a new symmetry, namely, the invariance under the rescaling of the frame. This last symmetry shows that the application of the converse of Noether's second theorem on particular models of $\fr$ gravity with torsion may lead to further symmetries.

\section{Symmetries of four--dimensional (4D) general relativity} \label{sec:4d}

We begin this section by recalling some facts concerning the symmetries of 4D general relativity in the Cartan formalism, and then we review the derivation of the internal gauge symmetry obtained in Ref.~\cite{Montesinos2017} for this case. This allows us to illustrate the basic idea behind the procedure that we use to uncover the analog of this gauge symmetry for $\fr$ gravity with torsion.   

Let $\mathcal{M}^4$ be a 4D orientable manifold and let $SO(4)$ be its frame rotation group for the Euclidean case $(\sigma=1)$ or $SO(3,1)$ for the Lorentzian one $(\sigma=-1)$\footnote{From here, we use the word `Lorentz' or `Lorentzian' for referring to both signatures, the Euclidean and the Lorentzian one.}; in each case, the associated metric is $(\eta_{IJ}):= \mathrm{diag}(\sigma,1,1,1)$. In the Cartan formalism, 4D general relativity  with cosmological constant $\Lambda$ is described by the Einstein--Cartan action (or Palatini action) $S[e,\omega]= \int_{\mathcal{M}^4} L_{\mathrm{GR}}$, whose Lagrangian $4$--form in terms of the orthonormal frame of $1$--forms $e^I$ and the spacetime connection $\omega^{I}{}_{J}$ compatible with the metric $\eta_{IJ}$, $d\eta_{IJ} - \omega^{K}{}_{I} \eta_{KJ} - \omega^{K}{}_{J} \eta_{IK} =0$ (and thus $\omega_{IJ}=-\omega_{JI}$), is given by 
\begin{eqnarray}\label{eq:grlag}
L_{\mathrm{GR}}=\frac{\kappa}{2}\epsilon_{IJKL} e^{I} \wedge e^{J} \wedge \left( R^{KL} - \frac{\Lambda}{3!}e^{K} \wedge e^{L} \right).
\end{eqnarray}
Here,  $R^{I}{}_{J}= d \omega^{I}{}_{J} + \omega^{I}{}_{K} \wedge \omega^{K}{}_{J}$ is the curvature of $\omega^{I}{}_{J}$ and $\kappa$ is a constant related to Newton's constant. The totally antisymmetric tensor $\epsilon_{IJKL}$ is such that $\epsilon_{0123}=1$ and the frame indices $I,J,K,\dots$ are raised and lowered with the metric $\eta_{IJ}$. The variational derivatives of the action defined by Eq.~(\ref{eq:grlag}) with respect $e^{I}$ and $\omega^{IJ}$ are
\numparts
\begin{eqnarray}
\fl \mathcal{E}_{I}&:= \frac{\delta S}{\delta e^{I}} =-\kappa \epsilon_{IJKL} \left(R^{JK}-\frac{\Lambda}{3}e^{J} \wedge e^{K} \right) \wedge e^{L}, \label{eq:eulerepal4d}\\
\fl \mathcal{E}_{IJ}&:= \frac{\delta S}{\delta \omega^{IJ}} =-\frac{1}{2} \kappa \epsilon_{IJKL} D \left( e^{K} \wedge e^{L} \right) \nonumber \\
\fl &\ = -\frac{1}{2} \kappa \epsilon_{IJKL} \left[ d \left( e^{K} \wedge e^{L} \right) + \omega^{K}{}_{M} \wedge  e^{M} \wedge e^{L} + \omega^{L}{}_{M} \wedge  e^{K} \wedge e^{M} \right], \label{eq:eulerwpal4d} 
\end{eqnarray}
\endnumparts
respectively, where $D$ is the covariant derivative defined by $\omega^{I}{}_{J}$~\cite{Castillobook}. Einstein's equations with cosmological constant follow from \eref{eq:eulerepal4d} and \eref{eq:eulerwpal4d}  by setting ${\mathcal E}_I=0$ and ${\mathcal E}_{IJ}=0$. Notice that if ${\mathcal E}_{IJ}=0$, then  the connection $\omega^{I}{}_{J}$ is torsion--free provided that the frame is nondegenerate. Nevertheless, we point out that throughout this paper the variational derivatives ${\mathcal E}_I$ and ${\mathcal E}_{IJ}$ will be assumed to be nonvanishing in general, since our approach to uncover gauge symmetries is {\it off--shell}.

Now we turn our attention to the gauge symmetries of the Einstein--Cartan action. The full gauge invariance of the action defined by Eq.~(\ref{eq:grlag}) can be equivalently described by two different sets of fundamental symmetries. The first set is composed of (infinitesimal) local Lorentz transformations
\begin{eqnarray}
\delta_{\tau}e^I =\tau^I{}_{J} e^J, \nonumber\\
\delta_{\tau}\omega^{IJ}=- D \tau^{IJ}=-\left(d\tau^{IJ}+\omega^I{}_K \tau^{KJ} +\omega^J{}_K \tau^{IK} \right),\label{eq:varwLorentz}
\end{eqnarray}
and (infinitesimal) diffeomorphisms
\begin{eqnarray}
\delta_{\xi} e^I = \mathcal{L}_{\xi}e^{I}, \nonumber\\
\delta_{\xi} \omega^{IJ} =\mathcal{L}_{\xi}\omega^{IJ}, \label{eq:varwdiffeos}
\end{eqnarray}
where the functions $\tau^{IJ}(=-\tau^{JI})$ are gauge parameters and ${\mathcal L}_{\xi}$ is the Lie derivative along the vector field $\xi$, which is the generator of a diffeomorphism. At this point it is worth recalling that an infinitesimal transformation of the fields depending on arbitrary functions is a gauge symmetry of the action if the corresponding Lagrangian remains quasi--invariant (invariant up to a total derivative) under it. In this regard, the change of the Lagrangian~(\ref{eq:grlag}) under local Lorentz transformations is $\delta_{\tau}L_{\mathrm{GR}}=0$ and under diffeomorphisms is $\delta_{\xi} L_{\mathrm{GR}} = d\left( \xi \intprod L_{\mathrm{GR}} \right)$, where $\xi \intprod L_{\mathrm{GR}}$ is the contraction of the vector field $\xi$ and the $4$--form $L_{\mathrm{GR}}$.

The second set of symmetries of the Einstein--Cartan action defined by Eq.~\eref{eq:grlag} is composed of (infinitesimal) local Lorentz transformations~(\ref{eq:varwLorentz}) and the internal gauge symmetry~\cite{Montesinos2017} 
\begin{eqnarray}
 \delta_{\rho}e^{I} &= D\rho^{I},\nonumber\\
 \delta_{\rho}\omega^{IJ} &=\frac{\sigma}{2} \left ( - \epsilon^{IJKL} \ast {\mathcal R}_{MKLN} + \ast {\mathcal R}\ast _{MN}\,^{IJ} \right )
\rho^M e^N + \Lambda \rho^{[I} e^{J]},\label{eq:wNewGR}
\end{eqnarray}
where we have written the curvature as $R^{I}{}_{J} = (1/2)\mathcal{R}^{I}{}_{JKL}e^{K} \wedge e^{L}$ and defined the left and right internal duals ${\ast {\mathcal R}}_{IJKL}:= (1/2) \epsilon_{IJ}\,^{MN} {\mathcal R}_{MNKL}$ and ${{\mathcal R}\ast}_{IJKL}:= (1/2) \epsilon_{KL}\,^{MN} {\mathcal R}_{IJMN}$, respectively. Also, $\rho^I$ is the gauge parameter associated to this transformation. It can be checked that the Lagrangian~(\ref{eq:grlag}) is quasi-invariant under the symmetry~(\ref{eq:wNewGR}), since
\begin{eqnarray}
\delta_{\rho} L_{\mathrm{GR}} = d\left[ \frac{\kappa}{2} \epsilon_{IJKL}  \rho^I \left(R^{JK} + \frac{\Lambda}{3} e^J \wedge e^K \right) \wedge e^L \right].
\end{eqnarray}
The internal gauge symmetry~\eref{eq:wNewGR} is the particular case for $n=4$ of the symmetry found in Ref.~\cite{Montesinos2017} for the $n$--dimensional Einstein--Cartan action, which corresponds to the higher--dimensional generalization of three--dimensional local translations~\cite{Witten1988,Carlip2003} (see~\cite{MontesinosMatter2018} for a nice derivation of this symmetry using the converse of Noether's second theorem). 

Since the set composed of local Lorentz transformations and the internal gauge symmetry~\eref{eq:wNewGR} is a ``complete set" (see Ref.~\cite{Henneauxbook}), it is possible to write an infinitesimal diffeomorphism acting on both the frame and the connection in terms of the symmetries of this set. As matter of fact, using the Cartan formula ${\mathcal L}_{X}Q = d(X \intprod Q) +X \intprod dQ$ with $Q$ being an arbitrary $k$-form, we can express Eq.~\eref{eq:varwdiffeos} as
\begin{eqnarray}
\delta_{\xi} e^I = & \left( \delta_{\rho} - \delta_{\tau}\right) e^{I} +\mbox{terms proportional to ${\mathcal E}_{IJ}$},\nonumber \\
\delta_{\xi}\omega^{IJ} &= \left( \delta_{\rho} - \delta_{\tau}\right) \omega^{IJ}+\mbox{terms proportional to ${\mathcal E}_{I}$},\label{eq:diffeostoLorentzneww}
\end{eqnarray}
where $\tau^{IJ} := \xi \intprod \omega^{IJ}$ and $ \rho^I:=\xi \intprod e^I$ are the field--dependent gauge parameters. This shows that, in this setting, infinitesimal diffeomorphisms can be regarded as a derived symmetry. The terms proportional to $\mathcal{E}_{I}$ and $\mathcal{E}_{IJ}$ in~\eref{eq:diffeostoLorentzneww} are known as ``trivial gauge transformations''. 

Let us now show how we can arrive at the internal gauge symmetry~\eref{eq:wNewGR} by using the converse of Noether's second theorem \cite{Noether1918,Noether1918Eng,Bessel-Hagen1921}, which is the fundamental tool that we use throughout this paper to uncover gauge symmetries. The converse of Noether's second theorem states that for every set of $m$ differential relations (Noether identities) among the variational derivatives of an action principle, corresponds a gauge symmetry involving $m$ gauge parameters. This means that we can replace the problem of finding infinitesimal gauge transformations that leave the action quasi--invariant by that of finding Noether identities.

With this in mind, the first step towards finding the internal gauge symmetry~\eref{eq:wNewGR} is to get the corresponding Noether identity. This is done as follows~\cite{Montesinos2017}. Taking the covariant derivative of Eq.~(\ref{eq:eulerepal4d}) and using the Bianchi identity $DR^{I}{}_{J}=0$, we obtain
\begin{eqnarray}
  D \mathcal{E}_{I}&= -\kappa \epsilon_{IJKL} \left(R^{JK} \wedge De^{L} - \Lambda De^{J} \wedge e^{K} \wedge e^{L}\right).\label{eq:covdereulere0}
\end{eqnarray}
Then, expressing the curvature as $R^{I}{}_{J} = (1/2)\mathcal{R}^{I}{}_{JKL}e^{K} \wedge e^{L}$ and using the fact that  $2De^{I} \wedge e^{J}\wedge e^K = D ( e^I \wedge e^{J}) \wedge e^{K} + e^{I} \wedge D(e^{J} \wedge e^{K}) - D ( e^I \wedge e^K) \wedge e^{J} $, Eq.~(\ref{eq:covdereulere0}) acquires the form
\begin{eqnarray}
  D \mathcal{E}_{I} &= -\frac{\kappa}{2} \epsilon_{IJKL}\mathcal{R}^{JK}{}_{PQ} \left[ D\left(e^{L} \wedge e^{P} \right)\wedge e^{Q} + \frac{1}{2} e^{L} \wedge D \left(e^{P}\wedge e^{Q}\right) \right]\nonumber\\
                    & + \frac{\kappa \Lambda}{2} \epsilon_{IJKL} D\left( e^{J} \wedge e^{K}\right) \wedge e^{L}.\label{eq:covdereulere1}
\end{eqnarray}
Now, Eq.~(\ref{eq:eulerwpal4d}) implies that $D\left(e^{I}\wedge e^{J}\right) = -(\sigma/2)\epsilon^{IJKL}\mathcal{E}_{KL}$, which substituted into Eq.~\eref{eq:covdereulere1} yields
\begin{eqnarray}
D\mathcal{E}_{I} - Z^{KL}{}_{IJ} e^{J} \wedge \mathcal{E}_{KL} = 0, \label{eq:NoetheridNewGR}
\end{eqnarray}
with 
\begin{equation}
 Z^{KL}{}_{IJ}:=\case{\sigma}{2}\left(-\epsilon^{KLMN}\ast\mathcal{R}_{IMNJ}+ \ast\mathcal{R}\ast{}_{IJ}{}^{KL}\right)+ \Lambda \delta^{[K}_{I} \delta^{L]}_{J},	\label{eq:ZtensorGR}
\end{equation}
where our convention for the antisymmetrizer is $A^{[IJ]}:=(A^{IJ}-A^{JI})/2$. This is the desired Noether identity. After multiplying Eq.~\eref{eq:NoetheridNewGR} by the gauge parameter $\rho^{I}$, we arrive at the \textit{off--shell} identity~\cite{Montesinos2017}
\begin{eqnarray}
	\mathcal{E}_{I} \wedge \underbrace{ D\rho^{I} }_{\delta_{\rho} e^{I}} + \mathcal{E}_{IJ} \wedge \underbrace{ Z^{IJ}{}_{KL} \rho^{K}e^{L} }_{\delta_{\rho}\omega^{IJ}} +  d(\rho^{I}\mathcal{E}_{I}) = 0.\label{eq:offshellNewGR}
\end{eqnarray}
From this, appealing to the converse of Noether's second theorem, we read off the internal gauge symmetry~\eref{eq:wNewGR} from the quantities multiplying $\mathcal{E}_{I}$ and $\mathcal{E}_{IJ}$ in Eq.~\eref{eq:offshellNewGR}.

\section{Symmetries of \texorpdfstring{$n$}{n}--dimensional \texorpdfstring{$\fr$}{fr} gravity with torsion}\label{sec:fr}

In order to make this paper self--contained, we will begin this section by giving a brief description of $n$--dimensional $\fr$ gravity with torsion in the Cartan formalism. We next focus on the main goal of the current paper, namely,  to uncover the $n$--dimensional analog of the internal gauge symmetry~(\ref{eq:wNewGR}) for $\fr$ gravity with torsion. To accomplish this, we will use the approach outlined in the previous section, which heavily relies on the use of the converse of Noether's second theorem.

Let $\mathcal{M}^n$ be an $n$--dimensional orientable manifold and let $SO(n)$ be the frame rotation group for the Euclidean case ($\sigma=1$) and $SO(n-1,1)$ for the Lorentzian one ($\sigma=-1$); to each case corresponds the metric $(\eta_{IJ}):= \mathrm{diag}(\sigma,1,\ldots,1)$. In the Cartan formalism, the action that describes $\fr$ gravity with torsion in $n$ dimensions ($n \geq 3$) is given by $S[e,\omega]=\int_{{\mathcal M}^n} L_{\fr}$, where the  Lagrangian $n$--form is~\cite{Rubilar_1998}
\begin{eqnarray}
L_{\fr} =\kappa f({\mathcal R}) \eta . \label{eq:frLagrangian}
\end{eqnarray}
Here, $\fr$ is an arbitrary (real) function of the Ricci scalar ${\mathcal R}:={\mathcal R}^{IJ}{}_{IJ}$ with $R^{I}{}_{J}:= d \omega^{I}{}_{J} + \omega^{I}{}_{K} \wedge \omega^{K}{}_{J}=(1/2){\mathcal R}^{I}{}_{JKL} e^K \wedge e^L$ the curvature of $\omega^{I}{}_{J}$, which is compatible with the metric $\eta_{IJ}$, $D\eta_{IJ} = d\eta_{IJ} - \omega^{K}{}_{I} \eta_{KJ} - \omega^{K}{}_{J} \eta_{IK} =0$ (and thus $\omega_{IJ}=-\omega_{JI}$), $e^{I}$ is an orthonormal frame of $1$--forms,  $\eta:=(1/n!)\epsilon_{I_{1} \ldots I_{n}} e^{I_{1}} \wedge \cdots \wedge e^{I_{n}}$ is the volume form, and $\kappa$ is a constant related to Newton's constant, whose numerical value depends on $n$. The frame indices $I,J,\ldots,$ now run from $0$ to $n-1$ and are raised and lowered with the metric $\eta_{IJ}$, and the totally antisymmetric tensor $\epsilon_{I_1 \ldots I_n}$ is such that $\epsilon_{0 \ldots n-1} =1$. 

The variational derivatives of the action defined by the Lagrangian $n$--form~\eref{eq:frLagrangian} with respect to the frame $e^I$ and the connection $\omega^{IJ}$ are, respectively:
\numparts
\begin{eqnarray}
\fl {\mathcal E}_I := \frac{\delta S}{\delta e^I} = \kappa(-1)^{n-1} \left[ f'({\mathcal R}) \star \left( e_I \wedge e_J \wedge e_K \right) \wedge R^{JK} + \left(\fr-{\mathcal R} f'{(\mathcal R})\right)\star e_I\right], \label{eq:eulere}\\
\fl {\mathcal E}_{IJ} := \frac{\delta S}{\delta \omega^{IJ}}=\kappa (-1)^{n-1} D\left[f'(\mathcal{R}) \star(e_I \wedge e_J) \right], \label{eq:eulerw}
\end{eqnarray}
\endnumparts
where $f'(\mathcal{R}):=\case{{\rmd}\fr}{\rmd\mathcal{R}}$ and `$\star$' is the Hodge dual operator:
\begin{equation}
\star(e_{I_1} \wedge \cdots \wedge e_{I_k})=\frac{1}{(n-k)!}\epsilon_{I_1 \ldots I_k I_{k+1} \ldots I_n} e^{I_{k+1}} \wedge \cdots \wedge e^{I_n}. \label{eq:Hodgedefinition}
\end{equation}
The equations of motion of the theory correspond to $\mathcal{E}_{I} = 0$ and $\mathcal{E}_{IJ} = 0$, which after some manipulations give rise
to
\numparts
\begin{eqnarray}
	f'({\mathcal R}) {\mathcal R}^{I}{}_{J} - \frac{1}{2}\fr \delta^{I}{}_{J} =  0, \label{eq:eulerefunctions}\\ T^{I}{}_{JK}=\frac{2}{\left(n-2\right) f'\left(\mathcal{R}\right)}\delta^{I}{}_{[J}\partial_{K]} f'\left(\mathcal{R}\right),\label{eq:eulerwfunctions}
\end{eqnarray}
\endnumparts
where ${\mathcal R}^I{}_J:={\mathcal R}^{KI}{}_{KJ}$ is the Ricci tensor, $\partial_{I}$ is the vector field dual to the frame $e^{I}$, {\it i.e.} $\partial_{J} \intprod e^{I}=\delta^{I}{}_{J}$, and  $T^{I}{}_{JK}$ are the components of the torsion 2--form $T^I=De^{I}= (1/2)T^{I}{}_{JK} e^{J} \wedge e^{K}$. Notice that with the particular choice $\fr= \mathcal{R}-2\Lambda$, Eq.~\eref{eq:eulerefunctions} leads to Einstein's equations with cosmological constant, whereas Eq.~(\ref{eq:eulerwfunctions}) implies that $T^{I}{}_{JK}=0$ and thus the connection $\omega^{I}{}_{J}$ is torsion--free. This is expected, since in this case the Lagrangian~\eref{eq:frLagrangian} reduces to the $n$--dimensional Einstein--Cartan Lagrangian with cosmological term,
\begin{eqnarray}\label{palatinindim}
L_{\mathrm{GR}}=\kappa \left(\mathcal{R}-2\Lambda\right) \eta = \kappa \left [ \star (e_I \wedge e_J) \wedge R^{IJ} - 2 \Lambda \eta \right ],
\end{eqnarray}
which for $n=4$ collapses to Eq.~(\ref{eq:grlag}). In the general case, from Eq.~(\ref{eq:eulerwfunctions}) it is seen that a non-linear function $\fr$ is the source of torsion, and hence the connection is no longer on-shell torsion-free even in vacuum. Furthermore, in contrast to general relativity, for a non--linear $\fr$ the right--hand side of Eq.~\eref{eq:eulerwfunctions} involves second derivatives of the connection. It should also be pointed out that as a result of the assumptions involved in the theories described by Eq.~\eref{eq:frLagrangian}, namely arbitrary torsion and vanishing non--metricity $D\eta_{IJ}=0$, the equation of motion~(\ref{eq:eulerwfunctions}) is different from its analogous counterpart in $\fr$ gravity in the Palatini formalism, where the assumptions are vanishing torsion and arbitrary non--metricity. Actually, in this last case, the corresponding equation of motion implies vanishing non--metricity only for a linear function $\fr$ (see Ref.~\cite{Hamity1993}, for instance).

Having introduced $\fr$ gravity with torsion, we now proceed to study its symmetries. We start by recalling that, by construction, the Lagrangian $n$--form~\eref{eq:frLagrangian} is invariant under local Lorentz transformations~\eref{eq:varwLorentz} and  quasi--invariant under infinitesimal diffeomorphisms~\eref{eq:varwdiffeos}. It is not hard to verify that the change of the Lagrangian $n$--form~\eref{eq:frLagrangian} under these symmetries is, respectively, $\delta_\tau L_{\fr}=0$ and $\delta_\xi L_{\fr}=d(\xi \intprod L _{\fr})$. Furthermore, it is well--known that local Lorentz transformations and diffeomorphisms can be used to capture the full gauge freedom of $\fr$ gravity with torsion.

Let us illustrate how the converse of Noether's second theorem can be used to obtain local Lorentz transformations of the $\fr$ Lagrangian \eref{eq:frLagrangian}. Taking the covariant derivative of Eq.~\eref{eq:eulerw}, we get
\begin{eqnarray}
\fl D {\mathcal E}_{IJ} &= (-1)^{n-1} \kappa D^2 \left[ f'({\mathcal R})\star\left(e_I \wedge e_J\right) \right] \nonumber\\
\fl & = (-1)^{n-1} \kappa f'({\mathcal R}) \left[ -R^{K}{}_{I} \wedge \star\left(e_K \wedge e_J\right)  -  R^{K}{}_{J} \wedge \star\left(e_I \wedge e_K\right) \right] \nonumber\\
\fl & = (-1)^{n-1}  \kappa f'({\mathcal R}) \left( - 2 {\mathcal R}^{K}{}_{[J} e_{I]} \wedge \star e_K \right),\label{eq:proofLorentz1}
\end{eqnarray}
where in the third line we have used the identity 
\begin{eqnarray}
\fl R^{KI} \wedge \star \left( e_J \wedge e_K \right) &= \frac{1}{n!}\epsilon_{K I_2\ldots I_n}{\mathcal R}^{K}{}_{J} e^{I} \wedge e^{I_2} \wedge \cdots \wedge e^{I_n} = - {\mathcal R}^{K}{}_{J} e^I \wedge \star e_K.
\end{eqnarray}
On the other hand, from Eq.~\eref{eq:eulere} we have
\begin{eqnarray}
e_I \wedge {\mathcal E}_J = \kappa (-1)^{n-1} \left( \fr e_I \wedge \star e_J -2 f'({\mathcal R}){\mathcal R}^{K}{}_{J } e_I  \wedge \star e_K \right).
\end{eqnarray}
Antisimmetrizing this expression  in the indices $I,J$  and inserting the result into Eq.~\eref {eq:proofLorentz1}, we arrive at the Noether identity
\begin{equation}
D {\mathcal E}_{IJ}-e_{[I}\wedge {\mathcal E}_{J]}=0,\label{eq:NoetheridLorentz2}
\end{equation}
which, after being multiplied by the gauge parameter $\tau^{IJ}(=-\tau^{JI})$ and some algebra, leads to the off--shell identity
\begin{eqnarray}\label{eq:NoetheridLorentz3}
{\mathcal E}_I \wedge \underbrace{\tau^I{}_{J} e^J }_{\delta_{\tau} e^I}+ {\mathcal E}_{IJ} \wedge \underbrace{  (- D \tau^{IJ})}_{\delta_{\tau} \omega^{IJ}} + d \left[(-1)^{n-1}\tau^{IJ} {\mathcal E}_{IJ}\right]=0.
\end{eqnarray}	
Appealing to the converse of Noether's second theorem, local Lorentz transformations~\eref{eq:varwLorentz} emerge from the quantities that multiply each variational derivative in Eq.~\eref{eq:NoetheridLorentz3}.

In addition to local Lorentz transformations and diffeomorphisms, we will show that the Lagrangian $n$--form~\eref{eq:frLagrangian} possesses a new internal gauge symmetry analogous to that of Eq.~\eref{eq:wNewGR}, and given by 
\begin{eqnarray}
\delta_{\rho} e^{I} = D \rho^I + Y_{n}{}^{I}{}_{JK} \rho^J e^K, \nonumber\\
\delta_{\rho} \omega^{IJ} = Z_n{}^{IJ}{}_{KL} \rho^K e^L, \label{eq:varwNew}
\end{eqnarray}
where
\numparts 
\begin{eqnarray}
\fl Y_{n}{}^{I}{}_{JK} & := & \frac{1}{(n-2)f'({\mathcal R})}\delta^{I}_{[J} \partial_{K]} f'({\mathcal R}), \label{eq:Ytensor}\\
\fl Z_{n}{}^{IJ}{}_{KL} & := & \frac{\sigma (n-3)}{(n-2)!}\left(\epsilon^{I J M I_1 \ldots I_{n-3}}\ast{}{\mathcal R}_{KI_1 \ldots I_{n-3}ML} +\ast{}{\mathcal R}\ast{}_{I_{1} \ldots I_{n-4}KL}{}^{I_{1} \ldots I_{n-4}IJ}\right)\nonumber\\
\fl & & + \frac{1}{(n-2)} \left( {\mathcal R}- \frac{\fr}{f'({\mathcal R})}\right)\delta^{[I}_K \delta^{J]}_L, \label{eq:Ztensor}
\end{eqnarray}
\endnumparts
with the $n$--dimensional left and right internal duals defined as
\numparts
\begin{eqnarray}
\ast{}\mathcal{R}_{I_1 \dots I_{n-2}MN} := \frac{1}{2}  \epsilon_{I_1 \dots I_{n-2}KL} \mathcal{R}^{KL}{}_{MN},\\
\mathcal{R}\ast{}^{MN I_1 \dots I_{n-2}} := \frac{1}{2} \epsilon^{I_1 \dots I_{n-2} KL} \mathcal{R}^{MN}{}_{KL},
\end{eqnarray}
\endnumparts
respectively. To prove this fact, we compute the change of the Lagrangian $n$--form~\eref{eq:frLagrangian} under the transformation~\eref{eq:varwNew}, obtaining
\begin{eqnarray}
\delta_\rho L_{\fr} = & d\left\{  \frac{\kappa}{(n-2)}\rho^I \star\left(e_I \wedge e_J \wedge e_K\right) \wedge \biggl[R^{JK} \biggr. \right. \nonumber\\
& \left.\left.+\frac{1}{(n-1)(n-2)} \left( {\mathcal R}-\frac{\fr}{f'({\mathcal R})}\right)e^J \wedge e^K\right] \right\}.\label{eq:varactionNew}
\end{eqnarray}
This means that the Lagrangian $n$--form~\eref{eq:frLagrangian} is quasi--invariant, and hence the transformation~\eref{eq:varwNew} is a gauge symmetry of the action principle built from~\eref{eq:frLagrangian}.

To uncover the internal gauge symmetry~(\ref{eq:varwNew}) we follow the procedure depicted in section \ref{sec:4d}, which in this case, involves constructing a Noether identity that relates the covariant derivative of ${\mathcal E}_I$ with both ${\mathcal E}_I$ and  ${\mathcal E}_{IJ}$. We start by computing the covariant derivative of Eq.~\eref{eq:eulere}, arriving at
\begin{eqnarray}
	\fl & & D{\mathcal E}_I  =  \kappa (-1)^{n-1} \left\{ f'({\mathcal R}) [ D \star\left(e_I \wedge e_J \wedge e_K\right)] \wedge R^{JK} \right. \nonumber \\
	\fl  & & + D f'({\mathcal R})\wedge \star \left( e_I \wedge e_J \wedge e_K\right)\wedge R^{JK}
	+ D\left( \fr-{\mathcal R} f'({\mathcal R})\right)\wedge \star e_I \nonumber\\
	\fl  & & \left. + \left( \fr-{\mathcal R} f'({\mathcal R})\right)\wedge D \star e_I \right\} \nonumber\\
	\fl &  & = \kappa (-1)^{n-1} \bigg\{ f'({\mathcal R}) \frac{(n-3)}{(n-2)!} \epsilon_{I J K I_1 I_{2} \ldots I_{n-3}} {\mathcal R}^{JK}{}_{MN}  \bigg. \nonumber\\
	\fl & & \times \left[
D \left(e^{I_1} \wedge e^{I_{2}} \wedge \cdots \wedge e^{I_{n-3}} \wedge e^M\right) \wedge e^{N} + \frac{1}{2} e^{I_1} \wedge D \left(e^{I_2} \wedge \cdots \wedge e^{I_{n-3}} \wedge e^M \wedge e^{N}\right)\right]  \nonumber \\
	 \fl & & + D f'({\mathcal R})\wedge \star \left( e_I \wedge e_J \wedge e_K\right)\wedge R^{JK}
	+D\left( \fr-{\mathcal R} f'({\mathcal R})\right) \wedge \star e_I \nonumber\\
	\fl & & \bigg. + \left( \fr-{\mathcal R} f'({\mathcal R})\right)\wedge D \star e_I \bigg\},\label{eq:NoetheridNewproc1}
\end{eqnarray}
where in the first equality we have used the Bianchi identity $DR^{IJ}=0$  whereas in the second the fact that
\begin{eqnarray}
\fl	(n-3)! \left[ D \star\left(e_I \wedge e_J \wedge e_K\right) \right] \wedge R^{JK} = \frac{(n-3)}{(n-2)} \epsilon_{IJK I_{1} I_{2} \cdots I_{n-3}}\mathcal{R}^{JK}{}_{MN} \nonumber\\
\fl \times \bigg[ D \left( e^{I_{1}} \wedge e^{I_{2}} \wedge \cdots \wedge e^{I_{n-3}} \wedge e^{M} \right) \wedge e^{N} + \frac{1}{2} e^{I_{1}} \wedge D\left(e^{I_{2}}\cdots \wedge e^{I_{n-3}} \wedge e^{M} \wedge e^{N}\right)\bigg],\label{eq:usefulDtoDD}
\end{eqnarray} 
which can be verified by a direct calculation. Then, the remaining task is to write the right--hand side of Eq.~(\ref{eq:NoetheridNewproc1}) in terms of ${\mathcal E}_I$ and  ${\mathcal E}_{IJ}$. This is done as follows.

Contracting Eq.~\eref{eq:eulerw} with $ \epsilon^{IJI_3 \ldots I_n}$, we have 
\begin{equation}
\fl f'({\mathcal R})D\left(e^{I_1} \wedge \cdots \wedge e^{I_{n-2}}\right) = \frac{\sigma(-1)^{n-1}}{2\kappa} \epsilon^{I J I_1 \ldots I_{n-2}}{\mathcal E}_{IJ} -Df'({\mathcal R}) \wedge \left(e^{I_1} \wedge \cdots \wedge e^{I_{n-2}}\right), \label{eq:eulerwtocovdern}
\end{equation}
which substituted into~Eq.~(\ref{eq:usefulDtoDD}) yields
\begin{eqnarray}
	\fl D{\mathcal E}_I & = & \frac{\sigma(n-2)}{(n-3)!} {\mathcal R}^{JK}{}_{MN}\epsilon_{IJK I_1 I_2 \ldots I_{n-3}}  \nonumber\\
	\fl & & \times \left( \epsilon^{I_1 I_2 \ldots I_{n-3} P Q M}{\mathcal E}_{PQ} \wedge e^N+ \frac{1}{2} e^{I_1} \wedge \epsilon^{I_2 \ldots I_{n-3} P Q M N}{\mathcal E}_{PQ}  \right) \nonumber\\
	\fl & & + \kappa (-1)^{n-1} \bigg[ \frac{1}{(n-2)}Df'({\mathcal R}) \wedge \star \left( e_I \wedge e_J \wedge e_K \right)\wedge R^{JK}\bigg.\nonumber\\
	\fl & & \bigg. +D\left( \fr-{\mathcal R} f'({\mathcal R})\right)\wedge \star e_I  + \left( \fr-{\mathcal R} f'({\mathcal R})\right)\wedge D \star e_I \bigg]. \label{eq:NoetheridNewproc2}
\end{eqnarray}
Now, from Eqs.~\eref{eq:eulere} and \eref{eq:eulerw} we obtain
\numparts
\begin{eqnarray}
	\star\left( e_I \wedge e_J \wedge e_K \right)\wedge R^{JK}  =\frac{(-1)^{n-1}}{\kappa f'({\mathcal R})}{\mathcal E}_I-\left(\frac{\fr}{f'({\mathcal R})} -{\mathcal R}\right)\star e_I,\\
	D\star e_I =\frac{(-1)^{n-1}}{\kappa f'({\mathcal R})(n-2)}e^J \wedge {\mathcal E}_{IJ}-\frac{(n-1)}{(n-2) f'({\mathcal R})}D f'({\mathcal R}) \wedge \star e_I,
\end{eqnarray}
\endnumparts
respectively. So, inserting these two expressions into Eq.~\eref{eq:NoetheridNewproc2}, we get
\begin{eqnarray}
		\fl D{\mathcal E}_I & = & \frac{\sigma(n-2)}{(n-3)!} {\mathcal R}^{JK}{}_{MN}\epsilon_{IJK I_1 I_2 \ldots I_{n-3}}  \nonumber\\
	\fl & & \times\left( \epsilon^{I_1 I_2 \ldots I_{n-3} PQ M}{\mathcal E}_{PQ} \wedge e^N+ \frac{1}{2} e^{I_1} \wedge \epsilon^{ I_2 \ldots I_{n-3} PQMN}{\mathcal E}_{PQ}  \right) \nonumber\\
	\fl & &  + \frac{1}{(n-2) f'({\mathcal R})}Df'({\mathcal R}) \wedge {\mathcal E}_I +\frac{1}{(n-2)}\left({\mathcal R-\frac{\fr}{f'({\mathcal R})}}\right)e^J \wedge {\mathcal E}_{IJ}\nonumber\\
	\fl & &  -\frac{\kappa(-1)^{n-1}}{(n-2)f'({\mathcal R})}\partial_I f'({\mathcal R})\left( n \fr -2{\mathcal R}f'({\mathcal R})\right)\eta  \label{eq:NoetheridNewproc3}.
\end{eqnarray}
The last term in Eq.~\eref{eq:NoetheridNewproc3} is rewritten using Eq.~\eref{eq:eulere} as
\begin{equation}
\kappa (-1)^{n-1} \left( n \fr -2{\mathcal R}f'({\mathcal R})\right) \eta = e^J \wedge{\mathcal E}_J,\label{eq:specialcase}
\end{equation}
and with this result, $D{\mathcal E}_I $ takes the final form 
\begin{eqnarray}
		\fl D{\mathcal E}_I  &=& -\frac{1}{(n-2) f'({\mathcal R})}\partial_I f'({\mathcal R})e^J \wedge {\mathcal E}_J+ \frac{\sigma(n-2)}{(n-3)!} {\mathcal R}^{JK}{}_{MN} 
		\epsilon_{IJK I_1 I_2 \ldots I_{n-3}}  \nonumber\\
	\fl && \times \left( \epsilon^{ I_1 I_2 \ldots I_{n-3} P Q M}{\mathcal E}_{PQ} \wedge e^N+ \frac{1}{2} e^{I_1} \wedge \epsilon^{ I_2 \ldots I_{n-3} PQMN}{\mathcal E}_{PQ}  \right) \nonumber\\
	\fl && + \frac{1}{(n-2) f'({\mathcal R})}Df'({\mathcal R}) \wedge {\mathcal E}_I  +\frac{1}{(n-2)}\left({\mathcal R-\frac{\fr}{f'({\mathcal R})}}\right)e^J \wedge {\mathcal E}_{IJ}. \label{eq:NoetheridNewproc4}
\end{eqnarray}

Substituting $Y_{n}{}^{K}{}_{IJ} $ and  $Z_{n}{}^{KL}{}_{IJ}$ given by Eqs.~\eref{eq:Ytensor} and \eref{eq:Ztensor} in Eq.~(\ref{eq:NoetheridNewproc4}), it is straightforward to arrive at the following Noether identity 
\begin{equation}
D {\mathcal E}_I - Z_n{}^{KL}{}_{IJ}e^J \wedge {\mathcal E}_{KL} - Y^{K}{}_{IJ} e^J \wedge {\mathcal E}_{K}= 0. \label{eq:NoetheridNew} 
\end{equation}
Multiplying  Eq.~\eref{eq:NoetheridNew} by the gauge parameter $\rho^I$ and after a bit of algebra, we get the {\it off--shell} identity
\begin{equation}
\fl {\mathcal E}_I \wedge \underbrace{ \left( D \rho^I + Y_{n}{}^{I}{}_{JK}\rho^J e^K \right)} _{\delta_\rho e^I} + {\mathcal E}_{IJ} \wedge \underbrace{Z_n{}^{IJ}{}_{KL}\rho^K e^L}_{\delta_\rho \omega^{IJ}}+d\left( (-1)^{n} \rho^I {\mathcal E}_I\right)=0. \label{eq:offshellNew}
\end{equation}
Taking into account the converse of Noether's second theorem, from the quantities multiplying ${\mathcal E}_I$ and ${\mathcal E}_{IJ}$ in Eq.~\eref{eq:offshellNew} we can read off a new gauge symmetry of the $\fr$ action, which is precisely that given in Eq.~\eref{eq:varwNew}.

An important remark about the internal gauge symmetry~(\ref{eq:varwNew}) is that it and diffeomorphisms are not independent gauge symmetries. To show this, it is convenient to write Eq.~\eref{eq:varwNew}  in the following alternative form
\begin{eqnarray}
\fl \delta_{\rho}e^I &= D\rho^I+\frac{f''({\mathcal R})}{(n-2) f'({\mathcal R})} \left(\partial_{K}{\mathcal R}\right)\rho^{[I} e^{K]}, \nonumber\\
\fl \delta_{\rho}\omega^{IJ} &= \left[C^{IJ}{}_{KL}-\frac{2(n-3)}{(n-2)} \delta^{[I}_{K}{\mathcal R}^{J]}{}_{L}\right]\rho^K e^L + \left[ \frac{2 {\mathcal R}}{(n-1)}-\frac{\fr}{(n-2) f'({\mathcal R})}\right]\rho^{[I} e^{J]}, \label{eq:varwNewalt}
\end{eqnarray}
where
\begin{eqnarray}
\fl	C_{IJKL} \equiv & {\mathcal R}_{IJKL}-\frac{1}{(n-2)}\left( \eta_{IK}{\mathcal R}_{JL}-\eta_{JK}{\mathcal R}_{IL}+\eta_{JL} {\mathcal R}_{IK}-\eta_{IL}{\mathcal R}_{JK}\right)\nonumber \\
\fl & +\frac{1}{(n-1)(n-2)} {\mathcal R} \left(\eta_{IK} \eta_{JL}-\eta_{IL} \eta_{JK} \right), \label{eq:Weyltensor}
\end{eqnarray} 
are the components of the Weyl tensor~\cite{Castillobook}.  Then, using Eq.~\eref{eq:varwNewalt} and the Cartan formula ${\mathcal L}_{X}Q = d(X \intprod Q) +X \intprod dQ$, we find that infinitesimal diffeomorphisms can be written as
\begin{eqnarray}
\fl \delta_{\xi} e^I =&  \left( \delta_{\rho} - \delta_{\tau}\right) e^{I} \nonumber +\frac{\sigma(-1)^{n-1}}{\kappa f'({\mathcal R})}\left[ \frac{2}{(n-2)}\star \left(e^K\wedge {\mathcal E}_{JK} \right)\rho^{[I} e^{J]} + \star\left(e^I\wedge {\mathcal E}_{JK} \right)\rho^{[J} e^{K]} \right], \nonumber\\
\fl	\delta_{\xi}\omega^{IJ}& = \left( \delta_{\rho} - \delta_{\tau}\right) \omega^{IJ}\nonumber\\
\fl &+\frac{\sigma(-1)^{n-1}}{\kappa f'({\mathcal R})}\left[ \frac{(n-3)}{(n-2)} \star\left( e^{[I} \wedge {\mathcal E}_K\right)\rho^{J]} e^K +\frac{3}{(n-2)} \star\left( e^{[I} \wedge {\mathcal E}_K\right)\rho^J e^{K]} \right],
\end{eqnarray}
where $\tau^{IJ} := \xi \intprod \omega^{IJ}$ and $ \rho^I:=\xi \intprod e^I$ are field--dependent gauge parameters and our convention for the antisymmetrizer is $A^{[IJK]}:=(A^{IJK}-A^{IKJ}+A^{JKI}-A^{JIK}+A^{KIJ}-A^{KJI})/3!$. This means that infinitesimal diffeomorphisms are linear combinations of local Lorentz transformations and  the transformation~\eref{eq:varwNew}, modulo terms proportional to ${\mathcal E}_I$ and  ${\mathcal E}_{IJ}$. This in turn implies that local Lorentz transformations and the new gauge symmetry can be taken as a fundamental set to describe the whole gauge symmetry of $\fr$ gravity with torsion.

Two comments are in order: (i) In the particular case of $n$--dimensional general relativity, namely $\fr=\mathcal{R}-2\Lambda$, the transformation~\eref{eq:varwNew} reduces to
\begin{eqnarray}\label{ngr}
\delta_{\rho} e^I =&& D\rho^I, \nonumber\\
\delta_{\rho} \omega^{IJ} =&& \frac{\sigma (n-3)}{(n-2)!} \Bigl( \epsilon^{IJLI_1 \dots I_{n-3}}\ast \mathcal{R}_{M I_1 \dots I_{n-3}LN} \nonumber\\
&& + \ast\mathcal{R}\ast _{I_1 \dots I_{n-4}MN}{}^{I_1 \dots I_{n-4}IJ} \Bigr) \rho^M e^N+ \frac{2 \Lambda}{n\!-\!2} \rho^{[I} e^{J]},
\end{eqnarray}
which is exactly the internal gauge symmetry reported in Ref.~\cite{Montesinos2017}. Furthermore, to obtain the transformation~(\ref{eq:wNewGR}) we only need to set $n=4$ in this expression, and to get three--dimensional local translations, 
\begin{eqnarray}
\delta_\rho e^I = D\rho^I,\qquad
\delta_\rho \omega^{IJ} =2\Lambda\rho^{[I} e^{J]},\label{eq:varw3dlt}
\end{eqnarray}
we simply set $n=3$ in Eq.~\eref{ngr}. Note that the second term in the gauge transformation of the frame $e^I$ given in Eq.~\eref{eq:varwNew} vanishes for the case of general relativity whereas if $\fr \neq \mathcal{R}-2\Lambda$ such a term will always be present.

(ii) It is important to point out that, in contrast to local Lorentz transformations and diffeomorphisms, the structure of the internal gauge symmetry~\eref{eq:varwNew} depends explicitly on the spacetime dimension $n$ and the function $\fr$ under consideration. For instance, in four dimensions $(n=4)$ the new symmetry~\eref{eq:varwNew} takes the form
\begin{eqnarray}
\fl \delta_{\rho}e^I = D\rho^I+\frac{1}{2 f'({\mathcal R})} \left( \partial_{J}  f'({\mathcal R}) \right)\rho^{[I} e^{J]},\nonumber\\
\fl \delta_{\rho}\omega^{IJ} = \frac{\sigma}{2}\left(-\epsilon^{IJKL}\ast{}{\mathcal R}_{MKLN} +\ast{}{\mathcal R}\ast{}_{MN}{}^{IJ}\right) \rho^{M}e^{N} + \frac{1}{2} \left( {\mathcal R}- \frac{\fr}{f'({\mathcal R})}\right)\rho^{[I} e^{J]},  \label{eq:varwNewalt4D} 
\end{eqnarray}
whereas in three dimensions $(n=3)$ the symmetry~\eref{eq:varwNew} becomes
\begin{eqnarray}
\delta_\rho e^I = D\rho^I + \frac{2}{f'({\mathcal R})} \left( \partial_J f'({\mathcal R}) \right) \rho^{[I} e^{J]},\nonumber\\
\delta_\rho \omega^{IJ} =\left({\mathcal R}-\frac{\fr}{ f'({\mathcal R})}\right)\rho^{[I} e^{J]}. \label{eq:varwNew3D}
\end{eqnarray}

Before concluding this section, we would like to remark that the converse of Noether's second theorem applied to some particular $\fr$ actions may lead to additional symmetries. As an example, notice that the left--hand side of Eq.~\eref{eq:specialcase} vanishes for $\fr = c{\mathcal R}^{n/2}$, with $c$ a real constant, which can be verified by a direct substitution. Then, in this case, we have the Noether identity
\begin{equation}
{\mathcal E}_I  \wedge e^I = 0. \label{eq:Noetheridsimple}
\end{equation}  
As for the previous Noether identities, we multiply Eq.~\eref{eq:Noetheridsimple} by the gauge parameter $\mu$, obtaining then the {\it off--shell} identity
\begin{equation}
{\mathcal E}_I \wedge \underbrace{\mu e^I}_{\delta e^I}= 0. \label{eq:off-shellsimple}
\end{equation}  
Once again, appealing to the converse of Noether's second theorem, we identify the gauge symmetry associated to the Noether identity~\eref{eq:Noetheridsimple} from the terms accompanying the variational derivatives in Eq.~\eref{eq:off-shellsimple}, namely
\begin{eqnarray}
\delta_{\mu}e^I = \mu e^I, \nonumber\\
\delta_{\mu}\omega^{IJ}= 0.\label{eq:varwsimple}
\end{eqnarray}
Hence, we can conclude that the action $S[e,\omega] = \kappa \int_{{\mathcal M}^n} {\mathcal R}^{n/2}\eta$ is invariant under the rescaling of the frame. Although it is well--known that the analogous action, in the Palatini formalism, $S[g,\Gamma]=\kappa \int \mathcal{R}^{n/2} \eta $, is invariant under conformal transformations of the metric~\cite{Ferraris1994}, the symmetry~\eref{eq:varwsimple} had not been reported in literature and can be considered as a new symmetry of the corresponding action. In this way, we have illustrated how new symmetries naturally emerge by applying our approach to particular cases of $\fr$ gravity with torsion.

\section{Conclusion}\label{Sec:Conclusion}

In this work, we have used the converse of Noether's second theorem to obtain a new internal gauge symmetry for $n$--dimensional $\fr$ gravity with torsion in the Cartan formalism that is the extension of the symmetry reported in Ref.~\cite{Montesinos2017} for general relativity. The new internal gauge symmetry has the following properties:
\begin{enumerate}
	\item The structure of the new internal gauge symmetry explicitly depends on the spacetime dimension $n$ and the particular form of the underlying $\fr$ function. Thus, the new symmetry depends on the dynamics of the theory, in contrast to local Lorentz transformations and diffeomorphisms which are insensitive to the form of the $\fr$ action.
	\item In the particular case of general relativity, that is $\fr = {\mathcal R}-2 \Lambda$, the  new internal gauge symmetry reduces {\it off--shell} to that of Ref.~\cite{Montesinos2017}, which is the  higher-dimensional generalization of 3D local translations. 
	\item The transformation of the frame $\delta_{\rho} e^{I}$ (see Eq.~(\ref{eq:varwNew})) involves an extra term as compared with its analog
	in the case of general relativity. Such a term vanishes when $\fr$ is a linear function of the Ricci scalar $\mathcal{R}$.
	\item The new internal gauge symmetry and local Lorentz transformations can be considered as a fundamental set of symmetries to describe the full gauge freedom of a general $\fr$ theory. This follows from the fact that infinitesimal diffeomorphisms can be written in terms of these symmetries. Then, in this framework, the whole gauge symmetry of $\fr$ gravity with torsion is purely internal and diffeomorphisms are no longer a fundamental symmetry.
\end{enumerate}

As future work, it would be interesting to obtain the finite gauge transformations corresponding to the symmetry \eref{eq:varwNew}, since this may have applications in the search for solutions of the field equations of $\fr$ gravity with torsion. On the other hand, due to its relevant role in the quantization of the theory, it would be desirable to obtain the gauge algebra of the new internal gauge symmetry (for a general $\fr$ theory with torsion) and local Lorentz transformations. This gauge algebra is expected to be generically open, just as in the case of general relativity~\cite{Montesinos2017}. Additionally, it is worth to explore if there exist functions $\fr$ for which the algebra of the new internal gauge symmetry and local Lorentz transformations closes. Finally, we would like to remark that analyzing particular cases of $\fr$ theories, or even other theories of gravity beyond general relativity, from the perspective of the converse of Noether's second theorem may be not only interesting, but also fundamental, since the last word about the ultimate nature of gravity has not been said yet, and so, this powerful mathematical tool may help us to get a deeper insight about it.

\ack
This work was partially supported by Fondo SEP--Cinvestav and by Consejo Nacional de Ciencia y Tecnolog\'ia
(CONACyT), M\'exico, Grant No. A1-S-7701. Diego Gonzalez is supported with a DGAPA-UNAM postdoctoral fellowship.
\section*{References}
\bibliography{Bibliography}
\end{document}